\documentclass[twocolumn,tighten,times,twocolappendix]{aastex631}

\graphicspath{{images/}}

\shortauthors{Bij et al.}

\begin{document}

\title{Kinematics of Crab Giant Pulses}

\author[0000-0001-7505-5223]{Akanksha Bij}
\affiliation{Canadian Institute of Theoretical Astrophysics,
  60 Saint George Street,
  Toronto, ON M5S 3H8, Canada}
\affiliation{Department for Physics, Engineering Physics and Astrophysics,
  Queen’s University,
  Kingston, ON, K7L 3N6, Canada}
\author[0000-0001-7453-4273]{Hsiu-Hsien Lin}
\affiliation{Canadian Institute of Theoretical Astrophysics,
  60 Saint George Street,
  Toronto, ON M5S 3H8, Canada}
\author[0000-0001-7931-0607]{Dongzi Li}
\affiliation{Canadian Institute of Theoretical Astrophysics,
  60 Saint George Street,
  Toronto, ON M5S 3H8, Canada}
\affiliation{Cahill Center for Astronomy and Astrophysics,
  California Institute of Technology,
  1216 E California Boulevard,
  Pasadena, CA 91125, USA}
\author[0000-0002-5830-8505]{Marten H.\ van Kerkwijk}
\affiliation{David A. Dunlap Department of Astronomy and Astrophysics,
  University of Toronto,
  50 Saint George Street,
  Toronto, ON M5S 3H4, Canada}
\author[0000-0003-2155-9578]{Ue-Li Pen}
\affiliation{Canadian Institute of Theoretical Astrophysics,
  60 Saint George Street,
  Toronto, ON M5S 3H8, Canada}
\affiliation{Department of Physics,
  University of Toronto,
  60 Saint George Street,
  Toronto, ON M5S 1A7, Canada}
\affiliation{Canadian Institute for Advanced Research,
  CIFAR Program in Cosmology and Gravity,
  Toronto, ON, M5G 1Z8, Canada}
\affiliation{Perimeter Institute for Theoretical Physics,
  Waterloo, Ontario N2L 2Y5, Canada}
\affiliation{Max-Planck-Institut f\"{u}r Radioastronomie,
  Auf dem H\"{u}gel 69,
  53121, Bonn, Germany}
\author[0000-0002-1568-7461]{Wenbin Lu}
\affiliation{Cahill Center for Astronomy and Astrophysics,
  California Institute of Technology,
  1216 E California Boulevard,
  Pasadena, CA 91125, USA}
\author[0000-0002-7164-9507]{Robert Main}
\affiliation{David A. Dunlap Department of Astronomy and Astrophysics,
  University of Toronto,
  50 Saint George Street,
  Toronto, ON M5S 3H4, Canada}
\affiliation{Max-Planck-Institut f\"{u}r Radioastronomie,
  Auf dem H\"{u}gel 69,
  53121, Bonn, Germany}
\author[0000-0003-1340-818X]{Jeffrey B.\ Peterson}
\affiliation{Department of Physics,
  Carnegie Mellon University,
  5000 Forbes Ave. Pittsburgh PA 15213 USA}
\author[0000-0002-7326-2779]{Brendan Quine}
\affiliation{Thoth Technology Inc.,
  33387 Highway 17,
  Deep River, Ontario K0J 1P0, Canada}
\affiliation{Department of Physics and Astronomy,
  York University,
  4700 Keele Street,
  Toronto, Ontario M3J 1P3, Canada}
\author[0000-0003-4535-9378]{Keith Vanderlinde}
\affiliation{David A. Dunlap Department of Astronomy and Astrophysics,
  University of Toronto,
  50 Saint George Street,
  Toronto, ON M5S 3H4, Canada}
\affiliation{Dunlap Institute for Astronomy and Astrophysics,
  University of Toronto,
  50 Saint George Street,
  Toronto, ON M5S 3H4, Canada}

\begin{abstract}
  The Crab Pulsar's radio emission is unusual, consisting predominantly of giant pulses, with durations of about a micro-second but structure down to the nano-second level, and brightness temperatures of up to $10^{37}\,$K.
  It is unclear how giant pulses are produced, but they likely originate near the pulsar's light cylinder, where corotating plasma approaches the speed of light.
  We report observations in the 400--800 MHz frequency band, where the pulses are broadened by scattering in the surrounding Crab nebula.
  We find that some pulse frequency spectra show strong bands, which vary during the scattering tail, in one case showing a smooth upward drift.
  While the banding may simply reflect interference between nano-second scale pulse components, the variation is surprising, as in the scattering tail the only difference is that the source is observed via slightly longer paths, bent by about an arcsecond in the nebula.
  The corresponding small change in viewing angle could nevertheless reproduce the observed drift by a change in Doppler shift, if the plasma that emitted the giant pulses moved highly relativistically, with a Lorentz factor $\gamma\sim10^4$ (and without much spread in $\gamma$).
  If so, this would support models that appeal to highly relativistic plasma to transform ambient magnetic structures to coherent GHz radio emission, be it for giant pulses or for potentially related sources, such as fast radio bursts.
\end{abstract}


\section{Crab Giant Pulses}
The Crab pulsar (PSR B0531+21) is a young pulsar, formed in supernova SN~1054, and powering a pulsar wind nebula in the Crab nebula supernova remnant.
It was the first out of eleven radio pulsars found to emit giant pulses \citep{staer68}.
Giant Pulses (GPs) are much shorter and brighter than regular pulses and occur in narrow phase ranges, typically limited to the edges of the main profile (for a description, e.g., \citealt{lyut07} and references therein).
In the case of the Crab, their intrinsic duration ranges from the order of nanoseconds to microseconds \citep{hank+03}, while their flux densities can reach brightness temperatures of $10^{37}\,$K \citep{hank+03,hanke07}.

The energetic GPs from Crab share numerous similarities with another type of energetic bursts, the Fast Radio Bursts (FRBs).
Both emit coherently with high polarization fractions;
both have short durations, with sub-structure down to even shorter timescales \citep{hank+03,nimm+21,maji21};
and while FRBs are brighter, the luminosities of the most energetic GPs approach those of the faintest FRBs \citep{hank+03,nimm+21,maji21}, and the gap is likely to be reduced as more nearby FRBs are found.
Given these similarities, it has been suggested that FRBs are GPs from (very) young, (very) rapidly rotating pulsars \citep{cordw16,lyu+21}.
Thus, constraints on the emission mechanism of GPs may also help elucidate the emission mechanism of FRBs.

Giant pulses likely originate near the pulsar's light cylinder, where corotating plasma approaches the speed of light \citep{eileh16}, but it is not yet clear what physical mechanism is responsible.
Indeed, as a result, theoretical estimates of the Lorentz factor $\gamma$ range from $\sim\!1$ to $\sim\!10^7$ (e.g., \citealt{eileh16, petr04, isto04, lyut07, lyub19, phil+19, mach+19, lyut21}).

For the Crab, unusually, the emission from giant pulses leads to the two largest components in the mean pulse profile: the main pulse and interpulse.
From detailed studies, \cite{hank+03, hanke07} found that there are two types of emission patterns.
The first type, in the main pulse and the interpulses at frequencies of up to 4 GHz, has micro-second duration bursts composed of nano-second duration shots.
The second, observed exclusively at higher frequency, between 5 and 30\,GHz, for a phase window near the interpulse (and hence called the high-frequency interpulse), has similar overall durations of a few microseconds, but has a spectrum showing proportionaly spaced emission bands.

In this work, we characterize Crab giant pulses at lower frequencies (400--800 MHz), where frequency-dependent scattering from the Crab Nebula becomes more pronounced, causing substantial smearing due to time delays from the emission travelling varying path lengths \citep{ranks73}.
Although we are unable to resolve the giant pulse `nanoshot' substructure at lower frequencies, the multiple lines of sight offer a unique opportunity to resolve the emission region \citep{main+21}, and, as we find here, to constrain properties of the emitting sources.

We describe our data in Section~\ref{sec:observations}, and the variety of giant pulse features we find in Section~\ref{sec:variety}, comparing with what was seen at higher frequency.
In Section~\ref{sec:origin}, we highlight spectral bands seen in some pulses, and how those drift in frequency within the scattering tail.
We interpret them in terms of a varying Doppler shift in Section~\ref{sec:model}, and explore alternative interpretations in Section~\ref{sec:alternatives}, before discussing implications in Section~\ref{sec:ramifications}.

\section{Observations and Data Reduction}
\label{sec:observations}
We observed the Crab pulsar with the 46m antenna at the Algonquin Radio Observatory (ARO) on 2015 July 23 and 24 (12.5 hours spread over 6 scans), and 2018 April 25 (1.5 hours over 2 scans).
Data were taken at 400-800 MHz with two linear polarizations.
The raw voltage streams (baseband) are processed using a CHIME acquisition board \citep{band+16}, which digitizes the signals, passes them through a polyphase filter, channelizes them to 1024 frequency channels, and then records these in standard VDIF format \citep{whit+09}.
Different receivers were used for the observations in 2015 and 2018, which had significantly different frequency responses and gains.
We gain and flux-calibrated these as described in Appendix~\ref{app:gains}.

The observation performed on 2015 July 24 partially overlaps with an observations of Crab pulsar at the John A. Galt Telescope at the Dominion Radio Astrophysical Observatory (DRAO).
These observations were originally taken for very long baseline interferometry with ARO, but here we will use them only to rule out instrumental effects at ARO.
They covered the same frequency band and were taken using a similar receiver and backend.

To search for giant pulses, we first coherently dedispersed the raw data to a reference frequency of 800\,MHz, using a dispersion measure (DM) of $56.7708{\rm\,pc/cm^{3}}$ for 2015 and $56.7562{\rm\,pc/cm^{3}}$ for 2018, obtained by using the closest DM value from the Jodrell Bank ephemeris \citep{lynepg93} as an initial guess and then optimized by adjusting the 4th significant figure such that the rising edge of the pulse is aligned, by-eye, in frequency.
Next, we sum the power from all frequency channels and both polarizations, apply a  rolling boxcar window of size 1.3\,ms, and flag any peaks more than 5$\sigma$ above the noise.
We calculate the rotational phase with the TEMPO2 package \citep{hobbem06}, using the Jodrell Bank ephemeris \citep{lynepg93}.
However, while our clocks were tied to a maser, they had small random offsets at the start of each integration.
Hence, while we can still distinguish main pulses from interpulses, otherwise only the relative phases are useful within each continuous integration.

For each pulse, we then created flux-calibrated dynamic spectra.

\begin{figure*}
  \centering
  \includegraphics[width=\hsize]{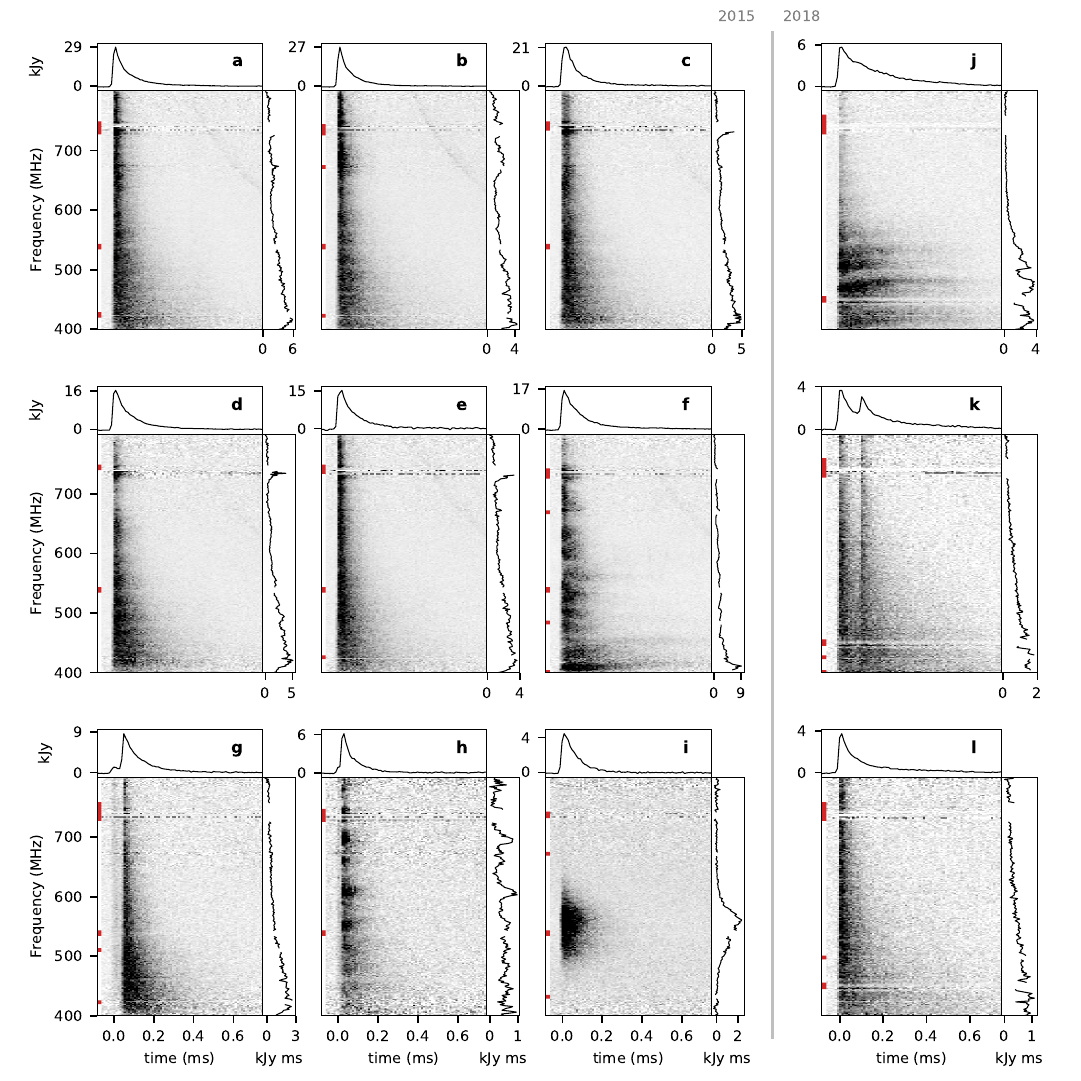}
  \caption{
    Dynamic spectra of 12 bright giant pulses.
    Each image shows flux-calibrated intensity over a 0.84\,ms time span and the full 400\,MHz frequency range, with the dedispersed data binned by 4 in both time and frequency (giving $10.24{\rm\,\mu s}$ time resolution and $1.56{\rm\,MHz}$ frequency resolution).
    The red marks to the left of the images indicate noisy frequency channels (determined from the channel standard deviation before gain correction).
    The top panels show frequency-averaged pulse profiles (in units of kJy), while the side panels show the pulse fluence spectra integrated over the full 0.84\,ms (in units of ${\rm kJy\,ms}$).
    Panels $a$--$f$ and $j$--$l$ are the brightest pulses in 2015 and 2018, respectively (ordered by descending brightness), while panels $g$--$i$ are three further 2015 pulses chosen for their particular profiles.
    Note that the faint curved feature visible in panels $a$--$f$, starting at $\sim\!0.4$ms at 800\,MHz and leaving the panels at $\sim\!650\,$MHz, is a reduction artefact (it represents incorrectly dedispersed leakage from neighbouring channels).
  }
  \label{figure:FigA}
\end{figure*}


\section{A Variety of Dynamic Spectra}
\label{sec:variety}
We show the resulting dynamic spectra for a selection of bright giant pulses in Figure~\ref{figure:FigA}.
This comparison readily reveals that there are large variations.
Some of the types we see can be related to what was seen previously, but others cannot.

In particular, from the figure, as well as from a categorization of the 148 brightest pulses by eye (see App.~\ref{app:categorization}), we find that most pulses are {\em`regular'} (e.g., panels $a$--$e$ and $l$ of Figure~\ref{figure:FigA}), showing only a smooth variation with frequency, and a scattering tail that becomes increasingly long towards lower frequency (with the scattering time longer in 2018 than in 2015).
In these, the scattering has likely blended together the short time-scale variations  seen at higher frequencies (1--43\,GHz; \citealt{hank+03,hanke07,eileh16}), where one often sees multiple micro-bursts over a few to a few tens of micro-second, with sub-structure down to the nano-second level.

Some other triggers (e.g., $g$ and $k$) are {\em`multi-peak'} pulses, with more than one obvious peak.
These likely reflect multiple indepedent giant pulses occurring in a single rotation, as also seen at higher frequencies \citep{eileh16}, but here partially blended together due to the scattering.

More unusal are the pulses that differ in their frequency spectrum.
Specifically, some show {\em`partial'} emission (e.g., $i$), where the pulse is bright in only a sub-band of frequencies rather than throughout the band, while
three others show {\em`banded'} spectra ($f$, $h$, and $j$), with the bands spaced by $\sim\!20$ to $\sim\!50{\rm\,MHz}$.
Possibly similar spectral structure at frequencies comparable to ours has been reported for giant pulses from the millisecond pulsars M28A \citep{bilo+15} and PSR~B1937+21 \citep{mcke+19}, but what is different for the three banded pulses we observe is that the bands appear to vary over the scattering tail, shifting in frequency or changing in relative strength.

Among the three banded pulses, $j$ is the clearest example, with the bands consistently moving up in frequency as the pulse proceeds.
Pulse $f$ also appear to have some drifting bands, but the direction is less clear: above 500\,MHz, there seems to be several bands that drift downward, while below 500\,MHz, some bands may be drifting up.

\section{Origin of the Spectral Bands}
\label{sec:origin}
At first glance, the drifting bands are reminiscent of a surprising earlier observation of the Crab, by \cite{hanke07}, who found that one particular pulse component, the high-frequency interpulse, showed bands with a typical spacing proportional to frequency,
$\Delta f\simeq0.06f$, which often drift upward over the few-$\rm\mu{}s$ duration of the pulse (see also \citealt{eileh16}).
It would be tempting to associate our banded pulses with these, especially as our observed spacing between bands, of $\Delta f\simeq40{\rm\,MHz}$ at $f\simeq500{\rm\,MHz}$, roughly follows the same scaling relation.
Nevertheless, the association seems unlikely, because the banded spectra we see are all from giant pulses associated with the Crab's main pulse component, for which banded spectra have never been seen at high frequencies.
In contrast the high-frequency interpulse is, as its name suggests, not visible at our relatively low frequencies.

A physical clue comes from the fact that the drifts and other variations in the banded spectra occur on the same timescale as the scattering, of a few ms, which is much longer than the duration of giant pulses.
Thus, the variation must be related to the fact that as time proceeds, one sees radiation from the pulse that has followed increasingly longer paths towards us, having been bent slightly by matter in the Crab nebula.

However, because the scattering timescale is much longer than the inverse of the band spacing, $\sim\!25{\rm\,ns}$, it is unlikely that the scattering itself causes the bands.
Given this, as well as the fact that the banding patterns are not shared between neighbouring pulses while at least for pulse $h$ they are seen in two telescopes simultaneously, we conclude that the banding is not due to instrumental or scattering effects, but intrinsic to the pulsar emission (for details, see App.~\ref{app:origins} and~\ref{app:scintillation}).
Indeed, banding itself seems not unexpected given the behaviour seen at high frequency, where giant pulses in the main pulse component often show several bright nano-shots spaced closely together.
If sometimes the broadband nano-shots are spaced by a few to a few tens of ns, this would at lower frequencies naturally lead to `partial' and `banded' spectra with structure on the scale of 10--100\,MHz.

\section{A Possible Physical Model}
\label{sec:model}

\begin{figure}
  \centering
  \includegraphics[width=\columnwidth]{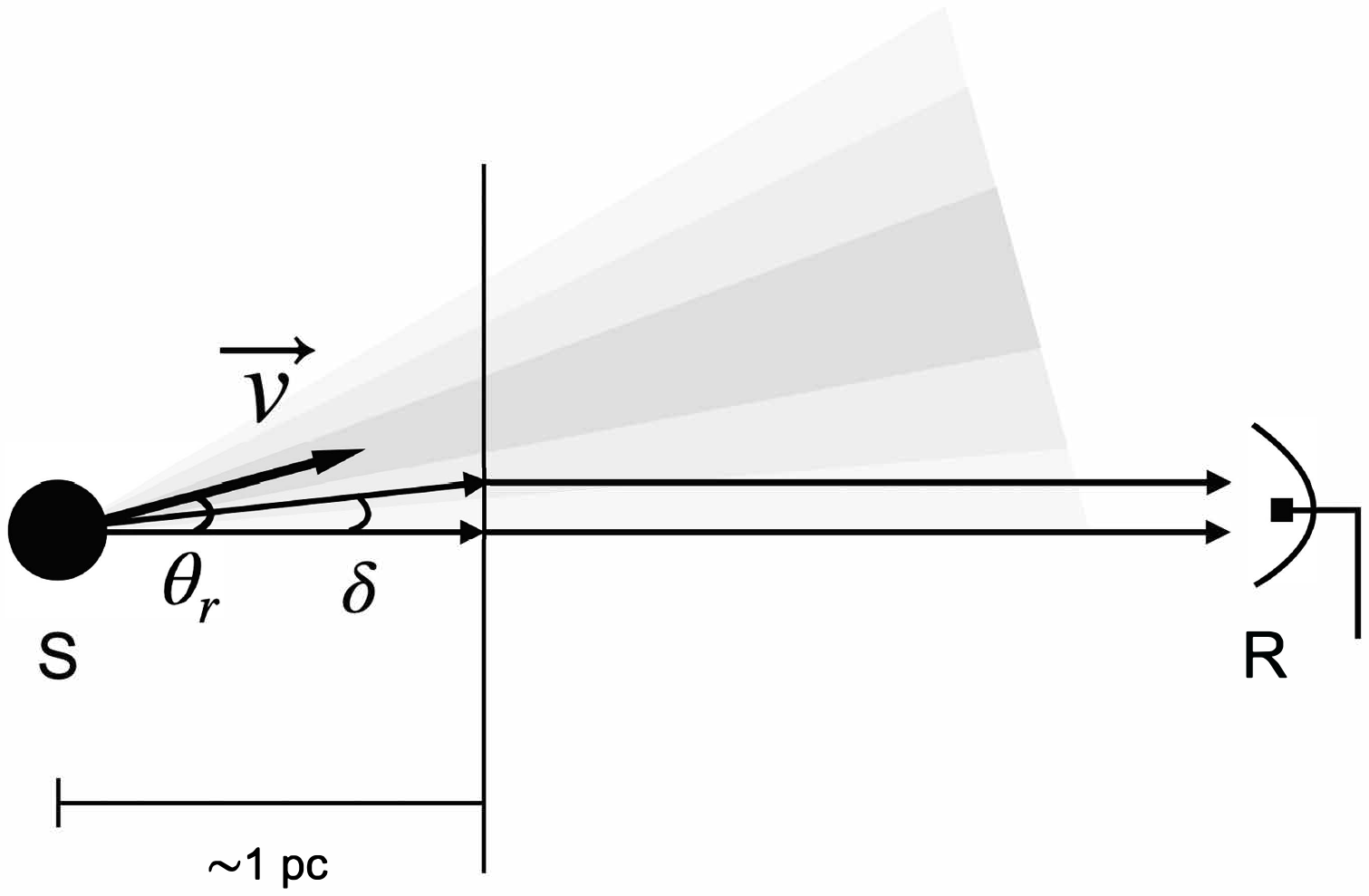}
  \caption{
    The model geometry: `S' is a blob of radiating plasma moving at an angle $\theta_{r}$ with respect to the observer `R', with a velocity $\vec{v}$ sufficiently relativistic that its emission is strongly beamed forward in the observer's frame (as indicated by the shading).
    At the start of the burst, the observer sees emission that travelled directly to the observer, but in the scattering tail the emission took a less direct path, initially leaving the source at an angle $\delta$ relative to the direct line of sight, before being bent towards the observer in the Crab Nebula (at about 1\,pc from the pulsar).
    If, as drawn, the angles are in the same direction, then as one proceeds along the scattering tail to larger $\delta$, one will see a steadily increasing blueshift.
  }
  \label{figure:geometry}
\end{figure}

If the banding in the spectra is indeed intrinsic to the emission, what could cause the changes during the scattering tail?
Inspired by the fact that in pulse $j$ the whole pattern moves in concert, we suggest that as the viewpoint shifts slightly during the scattering tail, one sees a difference in Doppler shift.
Specifically, we envision a situation like that depicted in Figure~\ref{figure:geometry}, where a group of charged particles moves relativistically, with a high Lorentz factor~$\gamma$.
Any emission from these particles will be beamed forward, within a beam of size $\sim\!1/\gamma$.
For that beam to be observable, it must include the line of sight, i.e., the particles should move at an angle $\theta_r\la1/\gamma$ relative to the line of sight.
The scattering screen will then bend light towards us which was originally emitted in a slightly different angle, either closer or further away from the direction the particles are moving in and thus with a different projected velocity.

Quantitatively, for the line of sight we expect emission to be Doppler shifted by a factor $D$ given by,
\begin{equation}
  D = \frac{1}{\gamma(1-\beta\cos\theta_{r})}
  \simeq \frac{2\gamma}{1+\gamma^2\theta_{r}^2}
\end{equation}
where $\beta = v/c$ and $\gamma = 1/\sqrt{1-\beta^2}$ as per convention.
The approximate equality is for highly relativistic motion, with $\gamma\gg1$, for which $\beta \simeq 1 - 1/2\gamma^2$ and $\cos\theta_{r} \simeq 1 - \frac{1}{2}\theta_{r}^2$ (the latter because $\theta_r$ has to be small, $\la1/\gamma$).

The above implies that for a fractional change in the viewing angle from $\theta_{r}$ to $\theta_{r}+\delta$, as resolved by our scattering screen, the resulting change in observed frequency $f_{r}$ will be,
\begin{equation}
  \frac{\Delta f_{r}}{f_{r}} = \frac{\Delta D}{D}
  = \frac{-\gamma^2(2\theta_{r}\delta+\delta^2)}{1+\gamma^2(\theta_{r}+\delta)^2}.
\end{equation}
For an observer in the center of the beam, with $\theta_{r}\ll\delta<1/\gamma$, one expects $\Delta f_{r}/f_{r} \approx -\gamma^2\delta^2$, indicating a downward drift. For the statistically more likely case depicted in Figure~\ref{figure:geometry}, with an observer at the edge of the beam, i.e., with $\theta_{r}\simeq1/\gamma$ , one finds $\Delta f_{r}/f_{r} \approx \gamma\delta$ (where the sign would be opposite if scattering was in the other direction, and cases where $\delta$ and $\theta_{r}$ are not aligned on the sky would be intermediate).

The scattering angle $\delta$ can be estimated using the known geometry for the Crab pulsar.
The distance of the screen is about $d\simeq1{\rm\,pc}$ from the pulsar \citep{95Lawr+,21Mart+}. The scattering time measured from the dynamic spectrum of the 2018 drifting pulse is approximately $\tau \simeq 0.5\,$ms at 450\,MHz.
Since from the geometry,
\begin{equation}
  c\tau = \frac{d}{\cos\delta} - d \simeq \frac{d}{2}\delta^{2},
  \label{sqrttau}
\end{equation}
we infer $\delta^2 \simeq 1\times10^{-11}$ and thus $\delta\simeq0.6\,$arcsec.
Given that we observe about a 4\% change in observed frequency, this implies $\gamma \simeq0.04/\delta\simeq10^4$.
Furthermore, given that we see the structure throughout the tail, the range in $\gamma$ cannot be very large,  $\Delta\gamma/\gamma\la 50\%$, i.e., the plasma must be relatively cold.
This is consistent with the presence of relatively narrow bands: we measure full width at half maximum $w\simeq20{\rm\,MHz}$, which would suggest a stricter constraint, $\Delta\gamma/\gamma\la w/f\simeq4\%$.

\begin{figure*}[!htb]
  \begin{center}
  \includegraphics[width=1.5\columnwidth]{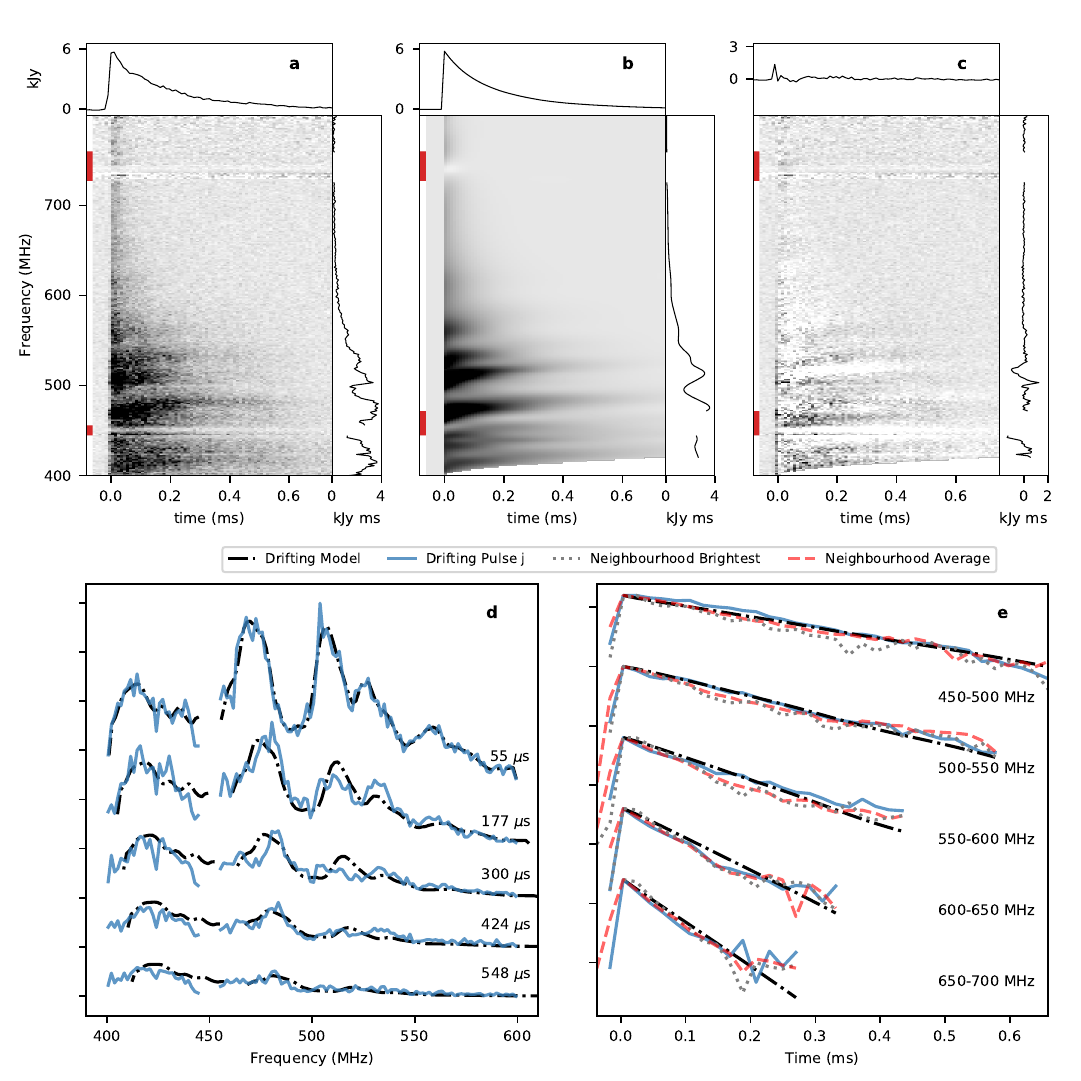}
  \caption{
    Comparison of pulse $j$ with the drifting bands with a simple model in which
    the initial profile is shifted in frequency and attenuated as time progresses.
    {\em Top row:\/} {\em a} The observed dynamic spectrum from Figure~\ref{figure:FigA}, {\em b} the approximate model, and {\em c} the residuals from the model.
    {\em d:\/} Spectra of pulse $j$ as a function of time, with the model overdrawn.
    {\em e:\/} Light curves throughout the scattering tail in various sub bands,
    with flux normalized to peak flux and shown in a logarithmic scale to ensure that exponential decay with time looks linear.
    The curves correspond to the drifting pulse $j$, the model, the next-brightest pulse within 5\,s and a stack of all pulses within 30\,s (see Fig.~\ref{figure:neighbours}).
    For the model, we started with the observed spectrum, a smoothed version of the average over the first 24 time samples ($61{\rm\,\mu s}$) of the pulse, and then shifted and dimmed it as described by Eq.~\ref{eq:model}.
    Given that scattering screen is unlikely to be as simple as assumed, the model reproduces the observations remarkably well.
    One also sees that pulse $j$ appears to dim slightly more slowly than the comparison pulses, as is expected in our model due to increasing Doppler boosting.
  }
  \label{figure:Fig6}
  \end{center}
\end{figure*}

If the above holds, then as a function of time $t$ relative to the start of the pulse, the Doppler shift should roughly vary as $\Delta D(t)/D \propto \delta(t) \propto \sqrt{t}$.
This scaling with the square root of time seems consistent from what one approximately sees for pulse $j$ in Fig.~\ref{figure:FigA}.

To explore this further, we compared the evolution of the observed spectrum with a simple model that starts with the initial spectrum but shifts and dims it over time,
assuming the $\sqrt{t}$ scaling above for the shifts and the standard exponential decay \citep{lang71} for the dimming, in which the decay timescale scales as $\tau\propto{}1/f^{4}$.
Specifically, we assume that the intensity $I(t, f)$ as a function of time since the start of the burst $t$ and frequency $f$ follows,
\begin{eqnarray}
  I(t, f) &=& I(0, f/D_{\rm rel}(t))\;D_{\rm rel}^{3}(t)\;{\rm e}^{-t/\tau(f)}\nonumber\\
  D_{\rm rel}(t) &=& D(t)/D(0) = 1+a\sqrt{t/\tau_{\rm ref}}\nonumber\\
  \tau(f) &=& \tau_{\rm ref}(f/f_{\rm ref})^{-4},
\label{eq:model}
\end{eqnarray}
where the term $D_{\rm rel}^{3}$ accounts for the fact that the Doppler effect not only shifts the frequencies, but also boosts the intensity due to beaming and time contraction.
For the initial spectrum $I(0, f)$, we use a smoothed version of the average over the first 24 time samples ($61{\rm\,\mu s}$) of the pulse.

In the expression for the decay time $\tau(f)$, either the reference frequency $f_{\rm ref}$ or the associated scattering time $\tau_{\rm ref}$ should be fixed.
Since we normalize the time in the Doppler shift term to $\tau_{\rm ref}$ to make the scale factor $a$ easier to interpret, we found it most convenient to fix the reference scattering time, to $\tau_{\rm ref}=250{\rm\,\mu{}s}$.
Then we roughly adjusted the two other parameters, finding that for  $a\simeq0.028$ and $f_{\rm ref}=465{\rm\,MHz}$ the model reproduces the observations quite well (see Fig.~\ref{figure:Fig6}), especially taking into account its simplicity.

In our scenario, different giant pulses would be emitted in different directions relative to the line of sight, and thus bands could drift up, down, or not at all.
Furthermore, for pulse $j$, we implicitly assumed that the scattering is asymmetric around the line of sight, dominated by the side in which pulse $j$ was emitted.
From observations of other pulsars, we know scattering geometries vary as the pulsar moves and probes different parts of the scattering screen \citep{hill+05}.
For the Crab, similarly, variations in scattering geometry on timescales of weeks are suggested by echoes and changes in scattering time \citep{mcke+18,drie+19}.
At times when the screens is more symmetric, one would expect to observe both upward and downward-drifting features simultaneously.
Since the structure of the scattering screen could be quite complex and the scattering strength is frequency dependent, an uneven distribution of material on different sides of the screen could result in the overall drift direction being frequency dependent.
Indeed, this may help understand the observed dynamic spectra of pulse $f$ (Fig.~\ref{figure:FigA}), which appears to have downward drifts at short delay, especially in the top half of the band and upwards shifts at longer delays in the bottom of the band.

An interesting aspect of our proposed Doppler-shift interpretation is the expectation that beyond the shift in frequency by a factor $D_{\rm rel}$ there will also be a boost in intensity by a factor $D_{\rm rel}^3$.
Since pulse~$j$ is drifting up, this boost factor will increase during the tail, and hence the scattering tail should be slightly brighter, by about 10\%, than would be the case for a pulse for which the Doppler factor did not vary.
To see if this is indeed the case, we also compare in Fig.~\ref{figure:Fig6} the brightness in the scattering tail of the drifting pulse $j$ with that of the brightest other pulse within 5\,s as well as that of the average of all pulses within 30\,s (see also App.~\ref{app:origins} and Fig.~\ref{figure:neighbours}).
One sees that the scattering tail of pulse $j$ does appear slightly boosted.
Furthermore, if one were to remove the Doppler boosting from the model, it would fall closer to the profiles of the other pulses.
However, the significance of the difference is limited, with the differences between pulse $j$ and the comparisons not much larger than those between pulse $j$ and the model.
A stronger test may be possible at lower frequency, where the range in scattering angle, and hence in Doppler shift, would be larger.

\section{Alternative Interpretations}
\label{sec:alternatives}

The drifts we observe in the banded structure of some pulses almost certainly reflect the change in viewing geometry that occurs during the scattering tail.
So far, we have assumed that the source of the burst intrinsically emitted a banded spectrum, and that the drift corresponded to a change in Doppler shift, from which we derived that the source had to move highly relativistically.

Here, we discuss what happens if one relaxes the assumption that the source produces bands intrinsically, instead requiring that the banding reflects interference between multiple nanoshots, which do not necessarily originate from the same region but are causally connected.
In this scenario, the banding in frequency comes from interference between multiple nanoshots, and the width of the bands thus reflects the inverse of the arrival time separation between different nanoshots.

In this picture, during the scattering tail, the nanoshots are viewed at an angle $\delta$ away from the direct line of sight, resulting in a change of path difference $\Delta x_\perp \delta$ between nanoshots, where $\Delta x_\perp$ is the distance between the nanoshots perpendicular to the line of sight (i.e., projected on the sky).
As the interference pattern of multiple nanoshots moves by about half of the pattern spacing in the scattering tail, $\Delta x_\perp \delta \simeq 0.5\lambda$.
Given our observation wavelength $\lambda\simeq0.5{\rm\,m}$ and $\delta \simeq 0.6{\rm\,arcsec}$, we infer a separation $\Delta x_\perp \sim 100{\rm\,km}$.

Given the typical observed frequency spacing between the bands of 
40\,MHz, in the observer's frame the nanoshots need to be separated in time by about 25\,ns.
Since at the speed of light this corresponds to only 7.5\,m, for the nanoshots to be causally connected, the signal must appear to move superluminally on the sky, by a factor of roughly $10^4$.
This is possible if the shots represent bursts arising from a single source that emits bursts (or causes bursts to be emitted) as it travels almost directly towards the observer at $\gamma\sim10^{4}$.

The above recovers the conclusions from Section~\ref{sec:model}, perhaps not surprising since neither mathematically nor physically can one distinguish between a blob moving relativistically that is emitting a banded spectrum or multiple nanoshots.
An implication in either case then is that the physical separation along the line of sight is a factor $\gamma$ larger than that on the sky, or a factor $\gamma^2$ relative to the observed time delay, i.e., about $10^8\times25\,\hbox{lt-ns}\sim2.5\,\hbox{lt-s}$ (or about 500 times the light cylinder radius $cP/2\pi\simeq1600{\rm\,km}$).


Instead of requiring the emission region of different nanoshots to be separated by $\sim100\,$km, alternatively, we could also consider nanoshots to be echos of a single impulse.
Here, the picture would be that the light emitted at an angle $\delta^\prime$ away from the line of sight encounters a large overdensity region of pulsar wind $d^\prime$ away from the emission region, which scatters it.
Similar to the calculation above, to explain the frequency widths of the bands, the scattered images, referred to as echoes, have to be delayed by $\tau^\prime=\frac{1}{2}d^\prime\delta^{\prime 2}/c\sim25{\rm\,ns}$.
In order for the nebular screen to resolve the different echoes in the scattering tail, they have to be separated by $d^\prime\delta^\prime \sim100{\rm\,km}$ during the first scattering. This yields $\delta^\prime \sim 10^{-4}{\rm\,rad} = 20{\rm\,arcsec}$ and $d^\prime\sim2.5\,\hbox{lt-s}$, i.e., also for this case something has to happen at a distance of about 500 times the light cylinder radius.

It is not obvious whether it is possible to produce such echoes in the pulsar wind, with the required rather large bending angles $\delta^\prime$, especially given the observation that giant pulses are composed of multiple nanoshots also at high frequency, where scattering would be less effective.
However, there are observational differences between this scenario and the Doppler shift scenario.
In particular, in this scenario, Doppler boosting does not play a role and hence one does not expect variations in the scattering tails of individual burst.
This unlike for the Doppler shift scenario, where especially at low frequencies one expects to occasionally observe a strongly boosted tail or a drop in flux corresponding to the scattered light coming from outside the relativistic beam.

\section{Ramifications}
\label{sec:ramifications}

We conclude that while our observations provide direct evidence that small changes in viewing direction during the scattering time result in different spectra, it is evidence only for highly relativistic motion if the shifts are interpreted as Doppler shifts.
In principle, it may be possible to get similar shifts in a model in which the banding results from interference of a few pulses that arise in different locations, as long as those locations are sufficiently far apart that their separation changes by 3\% as seen from the screen.
For all cases, however, this implies an extent of the region from which emission is seen by the screen of about $100{\rm\,km}$ as projected on the sky, and of about $2.5\,\hbox{lt-s}$ along the line of sight.

For the case that the plasma that emits the giant pulse moves highly relativistically, a prediction is that together with the frequency drift, there also is a Doppler boost in intensity, scaling as $\propto{}D^3$.
Thus, independent of any banding, pulses which drift upward should be slightly brighter in the scattering tail than the average, while those that drift downward should be fainter.
Unfortunately, in our data, the evidence for this effect is only marginal (see Sect.~\ref{sec:model} and Fig.~\ref{figure:Fig6}).


It should be possible to find more conclusive evidence at lower frequencies, where the scattering time is longer \citep{popo+06,karusl12,efte+16}, corresponding to larger changes in viewing geometry (one generically expects $\tau\propto1/f^{4}$ and thus $\delta\propto1/f^{2}$, which yields $\delta\simeq6{\rm\,arcsec}$ at 150\,MHz and $\delta\simeq40{\rm\,arcsec}$ at 60\,MHz).
Indeed, once the scattering angle $\delta$ becomes larger than the beam opening angle $1/\gamma$, one should sometimes see a suppression of the scattering tail.
For a given pulse, the exact effect would depend on geometry and the pulse's value of $\gamma$, but generically one predicts that individual pulses will no longer share the same scattering profile but rather show a range of behaviour around the average.

Taking our model results at face value, our inference of highly relativistic (but relatively cold) motion provides a strong constraint on the emission mechanism, since for different models vastly different values and distributions of $\gamma$ are expected, ranging anywhere from $\gamma\sim1$ to $\gamma\sim10^7$ (e.g., \citealt{eileh16, petr04, isto04, lyut07, lyub19, phil+19, mach+19, lyut21}).
A model that seems to match particularly well is the one recently proposed by \cite{lyut21}, in which the giant pulses are produced by pair plasma blobs created in reconnection events outside of the light cylinder, moving relativistically through ambient magnetic structures that causes the electrons and positrons to bunch up (and thus have a relatively narrow distribution in~$\gamma$) and emit coherent GHz
radio emission via the free electron laser mechanism \citep{colgn71}.
In this picture, for the parameters of the Crab pulsar, the plasma has to move with $\gamma\ga10^3$.
More generally, a large $\gamma$ value makes it easier to understand the high apparent brightness temperatures of the nanoshots.

The Crab's giant pulses are exceeded in brightness temperature only by the Fast Radio Bursts (FRBs).
Since FRBs share quite a number of properties with giant pulses, it may be that they also arise from plasma that moves highly relativistically \citep{lukz20}.
Repeating FRBs may be the closest analogues.
While those often show multiple burst substructure referred to as the ``downward march'' \citep{chime19} which given its persistent direction is unlikely to be related, it may still be worthwhile to look carefully at FRBs with long scattering tails.
Most promising may be those for which significant scattering has already been inferred, such as FRB~121102 \citep{jose+19}, FRB~181017.J1705+68 \citep{chime19}, and FRB~190117.J2207+17 \citep{fons+20} (though the first two are likely scattered in the milky way, making them less attractive candidates).
Also for the FRBs, high-time resolution observations at lower frequencies would be particularly valuable.

\begin{acknowledgments}
  A.~B. thanks Thierry Serafin Nadeau and Rebecca Lin for help with the data reduction.
  We also thank Laura Newburgh and Andre Renard for their critical contributions to the ARO 46m front-end and back-end systems, respectively, the latter and Tom Landecker for conducting the overlapping DRAO observations, Sasha Philippov, Maxim Lyutikov, Wadiasingh Zorawar and Sterl Phinney for discussions about the theoretical interpretation, all members of the scintillometry group for lively discussions, and the referee for comments that helped make our manuscript clearer.
  U.-L. P. receives support from Ontario Research Fund—Research Excellence Program (ORF-RE), Natural Sciences and Engineering Research Council of Canada (NSERC)  [funding reference number RGPIN-2019-067, CRD 523638-201, 555585-20], Canadian Institute for Advanced Research (CIFAR), Canadian Foundation for Innovation (CFI), Simons Foundation, Thoth Technology Inc, which owns and operates ARO, and Alexander von Humboldt Foundation.
  W.~L. was supported by the David and Ellen Lee Fellowship at Caltech.
  Computations were performed on the Niagara supercomputer at the SciNet HPC Consortium. SciNet is funded by: the Canada Foundation for Innovation; the Government of Ontario; Ontario Research Fund - Research Excellence; and the University of Toronto.
\end{acknowledgments}

\facility{Algonquin Radio Observatory:46m, DRAO:26m}
\software{astropy \citep{astropy18},
  Baseband (\citep{marten_van_kerkwijk_2020_4292543}),
  tempo2 \citep{hobbem06},
  numpy \citep{harr+20},
  matplotlib \citep{hunt07}}

\begin{appendix}
\section{Relative and Absolute Gain Calibration}
\label{app:gains}

\begin{figure*}
  \centering
  \includegraphics[width=2\columnwidth]{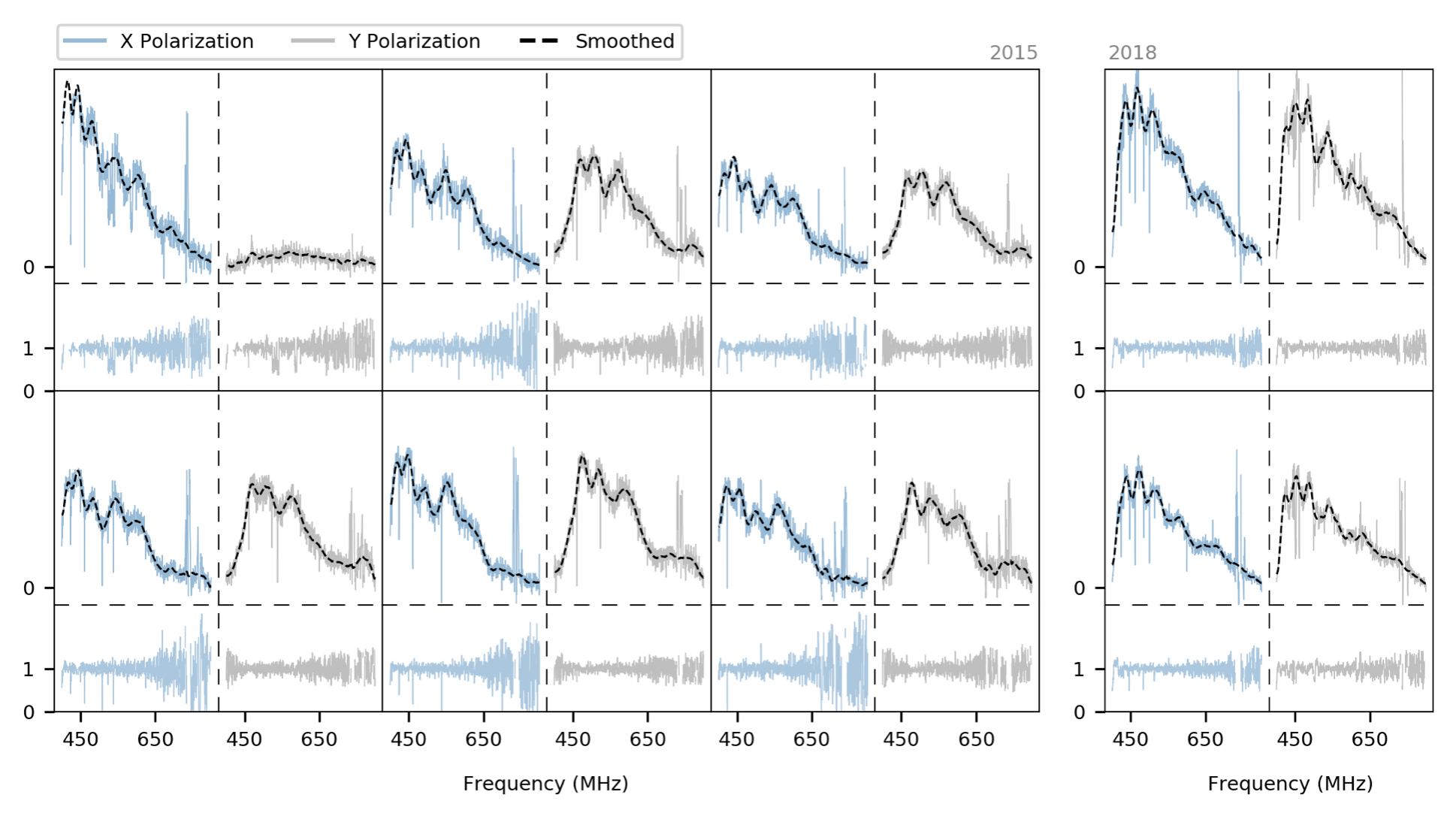}
  \caption{
    Gain determination using average spectra of giant pulses.
    For each of our 8 scans, the top panel shows the average background-subtracted giant-pulse spectra in X (\emph{left, blue}) and Y polarization (\emph{Right, gray}).
    Overdrawn are smoothed versions of the spectra (\emph{dotted, black}), which we use as a proxy for the instrumental gain.
    The bottom panels show the ratio between the raw and smoothed curves.
  }
  \label{figure:A1}
\end{figure*}

There were substantial differences between the frequency-dependent gains of the two polarizations, in particular in the 2015 data.
To help calibrate the differential gains, we calculated mean net fluxes in each polarization using bright pulses, using that the intrinsic polarization of giant pulses is quite random \citep{hankej16}, and hence an average should serve as an unpolarized source.
We choose the brightest 1000 pulses for each observating scan, subtracting off the off-pulse average, aligning and then averaging them to get the frequency profile.
We then use a smoothed profile as our relative gain, dividing all our intensities by them.

We show the result in Figure \ref{figure:A1}.
One sees that in the first section in 2015, the signal in Y was very weak.
Fortunately, that section contained none of the pulses of specific interest.
In the other 2015 sessions, the Y polarization shows a strong drop at the bottom of the band, while no such drop is seen in the X polarization.

To flux-calibrate the gain-corrected data, we use the Crab nebula itself, which contributes significantly to the off-pulse emission.
To get its net contribution, we subtract the average off-pulse signal from pointings towards other pulsars in the same run from the off-pulse signal from our pointings towards the Crab.
We then use that the nebular flux spectrum of the nebula is $F_{\rm neb}(f)\simeq955{\rm\,Jy}(f/{\rm1\,GHz})^{-0.27}$ (like \cite{cord+04}, we take the normalization from \cite{alle73} and slope from \cite{biet+97}), so that the flux of a giant pulse can be estimated from the measured (gain-corrected) intensities by,
\begin{equation}
F_{\rm GP}(f)=F_{\rm neb}(f)\,\frac{I_{\rm GP}(f)}{I_{\rm neb}(f)}.
\end{equation}

\begin{figure*}
  \centering
  \includegraphics[width=2\columnwidth]{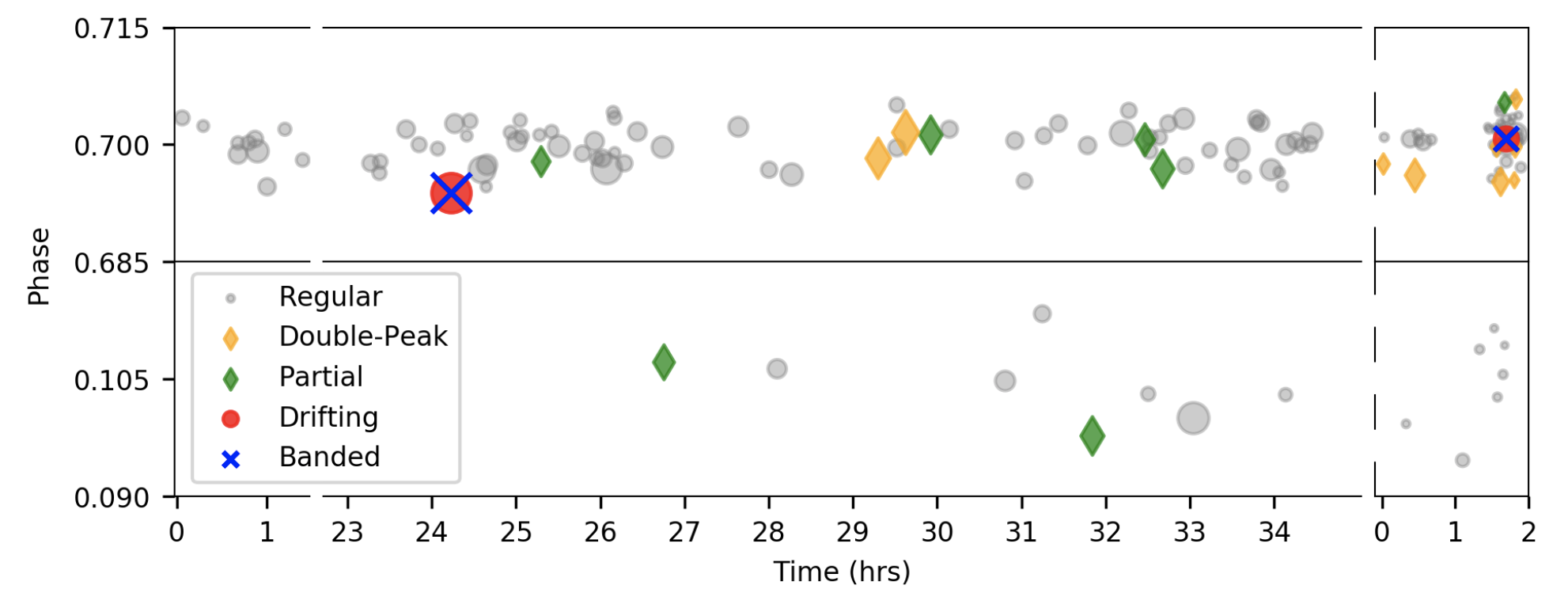}
  \caption{
    Categorized giant pulses as a function of time and phase.
    Marker styles reflect the different pulse categories (as labelled; see Table~\ref{tab:Table1}), while marker sizes scale with the peak flux.
    Breaks on the time axis are used to remove periods between scans, with the large break between the six scans in 2015 and the two scans in 2018 indicated by the dashed vertical line, and time on each side counted from the start of the first scan of each year.
    The top and bottom panels show phase windows around the main and interpulse components, respectively.
    Note that a constant phase offset has been applied in each scan to account for inaccuracies in the absolute clock times.
  }
  \label{figure:Fig2}
\end{figure*}

\begin{deluxetable}{lccc}
  \tablecaption{Giant pulse categorization.\label{tab:Table1}}
  \tablewidth{8cm}
  \tablehead{
    ~\hspace{2cm}~ && \colhead{Fraction} & \colhead{IP}\\[-.7em]
    \colhead{Feature} &\colhead{$N$} &\colhead{(\%)} &\colhead{(\%)}}
  \startdata
  All\dotfill & 148 & \llap{1}00 & 10\\[.5ex]
  Regular\dotfill & 129 & 87 & 10\\
  Multi-peak\dotfill & \phantom{00}9 & \phantom{0}6 & \phantom{0}0 \\
  Partial\dotfill & \phantom{00}7 & \phantom{0}5 & 30\\
  Banded\dotfill & \phantom{00}3 & \phantom{0}2 & \phantom{0}0 \\
  Drifting\tablenotemark{a}\dotfill & \phantom{00}2 & \phantom{0}1\rlap{.3} & \phantom{0}0 \\
  \enddata
  \tablenotetext{a}{`Drifting' is a sub-category of `Banded'.}
  \tablecomments{$N$ is the number of pulses that have a given feature, fraction the relative occurrence rate, and IP is the fraction that occurred in the interpulse phase.}
\end{deluxetable}

\section{Categorization of the Brighter Pulses}
\label{app:categorization}
We classified all 148 selected giant pulses by eye, using the categories given in Section~\ref{sec:variety}.
We summarize the results in Figure~\ref{figure:Fig2} and Table~\ref{tab:Table1}.
From Figure~\ref{figure:Fig2}, one sees that there are no obvious trends or clustering of pulse features with time and phase.

We note that all three banded pulses are very bright, with the drifting pulses $j$ and $f$ among the brightest of the pulses we detect.
This suggests our classification may be biased against faint pulses with banded structure.
It would be interesting to verify whether this is indeed the case, but we leave that to future work since here we are concerned with understanding the individual banded pulses themselves.

In contrast, numerous multi-peak pulses are found among the relatively fainter pulses.
This can be understood from the fact that in any given pair, at least one is likely to be faint, as faint pulses are much more common, and a faint additional pulse is easier to detect if the other pulse in the pair is faint as well.
Also here, we leave a more rigorous analysis for future work.

Among other bright pulses, $b$ and $d$ in Figure~\ref{figure:FigA} show hints of narrow bands, but the bands were too ambiguous for those pulses to be labelled as `banded'.
Similarly, while all the banded and multi-peak pulses are found among main-pulses, the low event-rate of interpulses does not allow us to rule out the possibility that these features also exist in the interpulse.

\section{Excluding Instrumental Origins}
\label{app:origins}

\begin{figure*}
  \centering
  \parbox[b]{0.65\hsize}{%
    \includegraphics[width=\hsize]{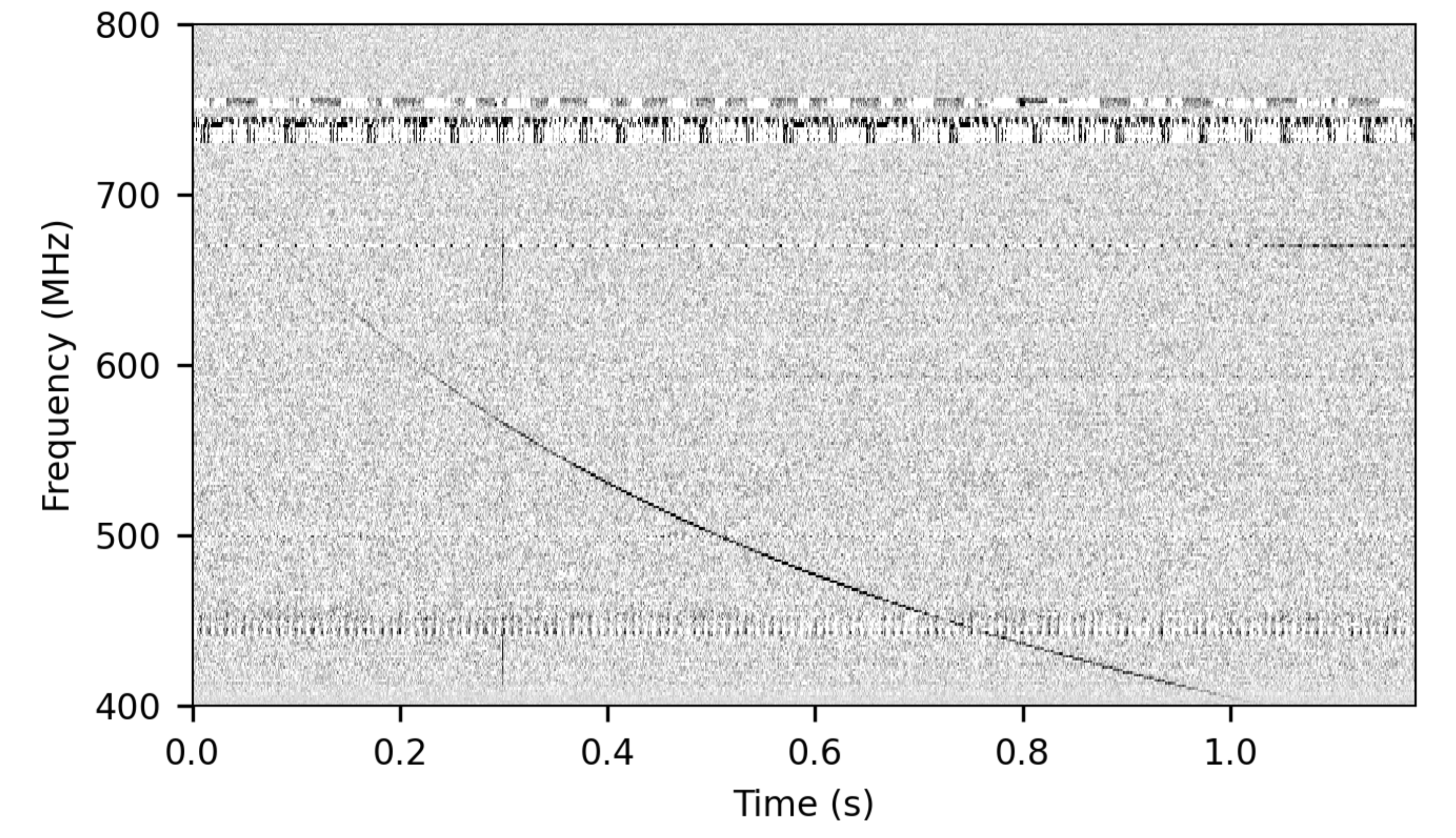}}\hfill
  \parbox[b]{0.3\hsize}{%
  \caption{Raw dynamic spectra of the giant pulse with drifting bands binned by 64 in time (163.84 $\mu$s) and 4 in frequency (1.5625 MHz).
    One sees that there are no drop outs or broad-band RFI events that could, after dedispersion, lead to bands in the frequency spectra.
    The structure around 475\,MHz is low-level radio-frequency interference, which makes that part of the spectrum somewhat less reliable.
  \label{figure:dispersed}
  }\vspace*{0.5cm}}
\end{figure*}

\begin{figure*}[t!]
  \centering
  \parbox[b]{0.65\hsize}{%
    \includegraphics[width=\hsize]{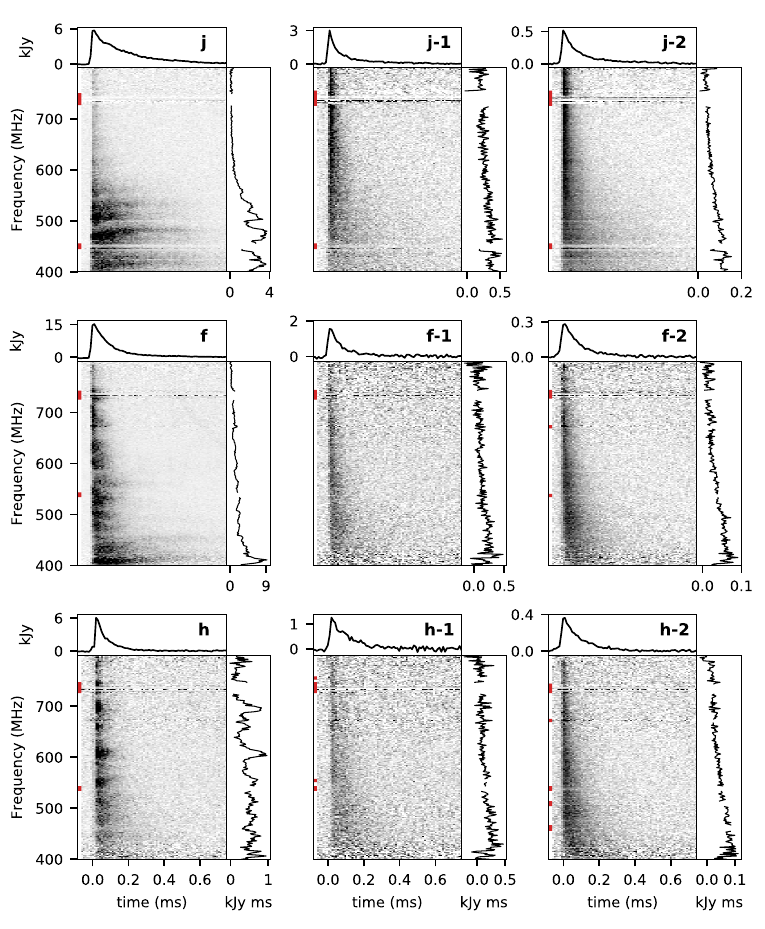}}\hfill
  \parbox[b]{0.3\hsize}{%
  \caption{
    Comparison of giant pulses showing banding in their spectra with neighbouring pulses.
   \emph{Left:\/} The three giant pulses with banding;
   \emph{Middle:\/} Next brightest single pulse within 5\,s of each banded pulse;
   \emph{Right:\/} Average of giant pulses in a 30\,s windows centered on the banded pulse.
   All spectra were constructed in the same way as those in Fig.~\ref{figure:FigA}.
   One sees that the banding does not persist beyond the banded pulses.
  \label{figure:neighbours}
  }\vspace*{0.5cm}}
\end{figure*}

\begin{figure}
  \centering
  \includegraphics[width=\columnwidth]{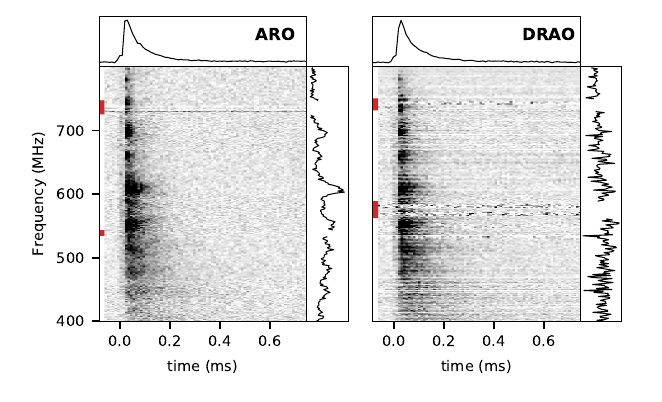}
  \caption{
    Dynamic spectra of banded giant pulse $h$ in Figure~\ref{figure:FigA} observed at two telescopes, ARO (\emph{Left}) and DRAO (\emph{Right}).
    The time and frequency resolution are the same as in Figure~\ref{figure:FigA}, and
    the data were reduced the same way, except without gain and flux calibration.
    While the DRAO data suffer from more artefacts (mostly because some channels are missing, causing the striated horizontal pattern), it is clear that they show the same frequency bands seen at ARO, thus confirming those are not instrumental in origin.
  }
  \label{figure:drao}
\end{figure}

In principle, it might be possible to produce apparent banding due to non-linearities, drop-outs, etc., in ways that might not be immediately obvious once data is dedispersed.
Hence, we looked in some detail at our data.

We started with possible problems with the data.  First, we ruled out that the banding and drifting effect is due to saturation, by verifying that the raw VDIF data for the time around the upper drifting pulse are not saturated:
We find that both the real and imaginary parts of the channel samples, of both polarizations and for all samples, have probability distributions that are approximately the same before, during, and after the sweep of the pulse (see Fig.~\ref{figure:dispersed}).
Other non-linear effects also seem unlikely, as the banding is also seen where the pulse is weaker and possible non-linear effect thus less strong: e.g., at high frequency for pulses $f$ and $h$ (where signals arrive before the brighter low-frequency signals due to dispersion), and in the scattering tail for all pulses (which arrives after the bright peaks).
We also looked for drop-outs in time, which in the dedispersed time stream might show up as bands\footnote{
  Any ``bands'' induced by time variability in the observed data would drift upwards slightly in the dedispersed data.
  Given the dispersion measure of $\sim\!57{\rm pc/cm^{3}}$ of the Crab, however, the expected drift rate of $\sim\!0.19{\rm\,MHz/ms}$ at $450\,$MHz is far smaller than the drift rates shown by some of our banded pulses.}.
We found no such drop-outs (see Fig.~\ref{figure:dispersed}).
Finally, since some giant pulses are strongly polarized and the intensity spectra might be affected by polarization leakage, we confirmed that the spacing of none of the bands matches that expected from the known Faraday rotation.

Beyond the individual time segments, we also checked for possible longer term problems with the instrumental response using data recorded close in time to each banded pulse.
In Figure~\ref{figure:neighbours}, we compare the dynamic spectra of the three banded pulses with the next brightest pulse near each (at most 5\,s away), and with the sum of all pulses near it (within a 30\,s window centered on the pulse).
One sees that the neighbouring pulses do not share the bands, indicating that the bands are intrinsic to each pulse.

The strongest evidence against an instrumental origin of the banded feature is that one banded pulse ($h$ in Fig.~\ref{figure:FigA}) is simultaneously observed at DRAO and ARO (while the ARO timestamp has some uncertainty due clocks not being synchronized with GPS, it is easily sufficient to identify a given Crab cycle).
In Fig.~\ref{figure:drao}, we show the burst as detected at the two telescopes
(after de-dispersion and subtraction of the off-pulse, but without gain correction).
One sees that the banding is similar in both.

\section{Excluding Interplanetary and Interstellar Scintillation}
\label{app:scintillation}

The light from the pulsar is bent due to electron density fluctuations in the surrounding Crab nebula, causing delays seen in the scattering tail.
On the way to us, the light also encounters electron density fluctuations in the interstellar medium and in the solar wind, which cause interstellar and interplanetary scintillation, respectively.



Interplanetary scintillation (IPS) generally causes only weak modulation in our frequency band, but since we observed an unusual event, we will nevertheless consider whether a strong IPS event could reproduce the drifting bands that we observed.
The typical timescale on which IPS modulates flux is $\Delta t\sim{\rm few~s}$.
Given a typical velocity in the solar wind of $v\sim500{\rm\,km/s}$, the size of the scattering involved is thus $L=v\Delta t \sim10^3{\rm\,km}$, which corresponds to scattering angles of $\theta = L/\lambda \sim 0.1{\rm\,arcsec}$ (for wavelength $\lambda=0.5{\rm\,m}$ [600~MHz]).
For a typical distance to a density perturbation $d\sim1{\rm\,AU}$, the expected scattering time is then $\tau=\frac{1}{2}d\theta^2/c\sim0.1\,$ns,
which corresponds to a de-correlation bandwith $\sim\!1/2\pi\tau\sim500{\rm\,MHz}$.
The latter is larger than our observation bandwidth and hence IPS would appear as a modulation in time, affecting all frequencies simultaneously.

Since the pulse is dispersed in the interstellar medium, the IPS time modulation would appear as bands in frequency after de-dispersion.
However, like the instrumental effects discussed above, the bands would not drift noticeably.
Moreover, due to the proximity of the source of interplanetary scintillation, the scintillation pattern seen by two telescopes separated by more than the size $L\sim10^{3}{\rm\,km}$ should not be the same, yet we detected one banded pulse simultaneously at ARO and DRAO (which are separated by about $3000{\rm\,km}$).
Thus, we conclude that the banding we observe is not introduced by IPS.

The Crab pulsar is known to be affected by interstellar scintillation.
Interpolating from the observed angular broadening of the Crab in Very Long Baseline Interferometry \citep{vand76,rudn+16}, we infer a scattering angle $\theta\sim1.5\,$mas at 600\,MHz for interstellar scintillation.
For a screen roughly halfway to the pulsar, at $d_{\rm screen}\simeq1{\rm\,kpc}$, this implies a scattering time $\tau=\frac{1}{2}d_{\rm eff}\theta^{2}/c\simeq5{\rm\,\mu s}$ (where the effective distance $d_{\rm eff} = d_{\rm screen}d_{\rm psr} / (d_{\rm psr}-d_{\rm screen})\simeq2\,$kpc).
Given this, the de-correlation bandwidth is $\sim\!1/2\pi \tau\simeq30{\rm\,kHz}$.
Since this value is much smaller than the $390\,$kHz width of our frequency channels, it would be unlikely to be observable, let alone able to explain the observed $\sim\!40{\rm\,MHz}$ band spacing.

In addition, no frequency modulation would be expected if the Crab pulsar, broadened by nebular scattering, is not a point source to the interstellar screen.
Since the physical size of the interstellar scattering region is roughy $L=\theta d_{\rm screen}\simeq1.5$\,AU, it will resolve any source at the distance of the Crab that is larger than $(\lambda/L)(d_{\rm psr}-d_{\rm screen})\simeq0.5\,$mAU.
This can be compared with the size of the nebular scattering region: as seen from the pulsar, we inferred from the observed scattering time that it subtends an angle $\delta \simeq 0.6{\rm\,arcsec}$ (see Sec.~\ref{sec:model}), which, at $\sim\!1{\rm\,pc}$ from the pulsar, implies a scattering disk size of $\sim\!0.6{\rm\,AU}$.
Thus, as was already noted by \cite{vand76}, the scattering region in the nebula is too large for there to be any scintillation due to the interstellar medium.


\end{appendix}

\bibliography{references}{}
\bibliographystyle{aasjournal}
\end{document}